\documentclass[aps,twocolumn,floats,prd,nofootinbib,10pt,longbibliography,superscriptaddress]{revtex4-2}

\usepackage[dvips]{graphicx} %
\usepackage{graphicx,amsmath,amsfonts,amssymb,slashed,hyperref}
\usepackage[normalem]{ulem}
\usepackage{bbold,wasysym}
\usepackage{epsfig}
\usepackage{array,multirow}
\usepackage[utf8]{inputenc}

\usepackage[usenames,dvipsnames]{xcolor} 
\usepackage{adjustbox}
\usepackage{soul}
\usepackage{orcidlink}

\newcommand{\class}{{\sc class}}

\begin{document}

\begin{flushright}
    \texttt{TTK-25-31}
\end{flushright}

\title{Constraints on neutrino mass and dark energy agnostic to the sound horizon}

\author{Ravi Kumar Sharma\orcidlink{0000-0002-8950-8446}}
\email{rksharma@physik.rwth-aachen.de}

\affiliation{Institute for Theoretical Particle Physics and Cosmology  (TTK),
RWTH Aachen University, Sommerfeldstr. 16, D-52056 Aachen, Germany}
\author{Julien Lesgourgues\orcidlink{0000-0001-7627-353X},}
\email{lesgourg@physik.rwth-aachen.de}
\affiliation{Institute for Theoretical Particle Physics and Cosmology  (TTK),
RWTH Aachen University, Sommerfeldstr. 16, D-52056 Aachen, Germany}

\begin{abstract} 
 Recent BAO observations from DESI DR2 either hint at a possible dynamical dark energy component, which would worsen the Hubble tension, or at a 95\% credible interval for the summed neutrino mass hardly compatible with neutrino oscillation experiments. In this context, it is interesting to investigate constraints on neutrino masses, dark energy and the Hubble parameter that are agnostic to some aspects of the cosmological model. Here we choose to be agnostic to the value of the sound horizon at recombination, while sticking to standard assumptions regarding the time of recombination and the growth of structures. To be consistent, we also disregard information on the full shape of the CMB temperature and polarization spectrum on sub-degree scale. With such agnostic and conservative assumptions, using data mainly on uncalibrated distances, the growth of structures, and laboratory bounds on tritium $\beta$-decay, we find that: (i) the dark energy evolution is well constrained by uncalibrated data on angular and luminosity distances, with a mild preference for dynamical dark energy, independently of the value of the sound horizon; (ii) large values of the Hubble rate are favored, 
$H_0=74.7^{+3.4}_{-4.4}$
 km/s/Mpc (68\%CL), together with low values of the sound horizon,
 $r_{\rm s}=131.1^{+6.8}_{-6.9}$
 Mpc (68\%CL); the SH0ES value of $H_0$ is thus marginally preferred over the low value returned by the standard inverse distance ladder analysis; (iii) the cosmological neutrino mass bound relaxes to
 $\sum m_\nu = 0.69^{+0.33}_{-0.47}$  eV (68\%CL)
 and becomes well compatible with the normal and inverted neutrino mass schemes.
\end{abstract}

\maketitle

\section{Introduction}
\label{sec:intro}

In the standard model of cosmology, $\Lambda$CDM, the two densest components in the present universe are the cosmological constant $\Lambda$ and cold dark matter (CDM). This model provides a successful fit to cosmological data on Cosmic Microwave Background (CMB) anisotropies \cite[e.g.][]{Efstathiou:1992sy,2013ApJS..208...20B,Planck:2018vyg,SPT-3G:2022hvq,ACT:2025fju}, Baryon Acoustic Oscillations (BAO) \cite[e.g.][]{2011MNRAS.416.3017B,Ross:2014qpa,BOSS:2016wmc,eBOSS:2019qwo,eBOSS:2019ytm,DESI:2024mwx,DESI:2025zgx}, the uncalibrated luminosity of type Ia supernovae (SNIa) \cite[e.g.][]{Brout:2022vxf,DES:2024jxu,Rubin:2023ovl}, and the large scale structure of the universe \cite{DES:2025tir}. However, a key challenge to the $\Lambda$CDM model is the Hubble tension: direct measurements of the current expansion rate give a high value, $H_0 = 73.17 \pm 0.86$ km/s/Mpc in the case of SH0ES \cite[e.g.][]{Riess:2021jrx,Breuval:2024lsv} (see also \cite{Freedman:2024eph}), while indirect measurements based on CMB and BAO data (assuming $\Lambda$CDM) yield a smaller value, $67.24 \pm 0.35$ km/s/Mpc in the case of recent combined CMB data \cite{SPT-3G:2025bzu} (see \cite[e.g.][]{Verde:2023lmm,Schoneberg:2021qvd} for reviews). Additionally, the most recent BAO measurements \cite[e.g.][]{DESI:2024mwx,DESI:2025zgx} suggest a possible 2 -3 $\sigma$ tension in the determination of the cosmological constant amplitude, parametrized in $\Lambda$CDM through the fractional density $\Omega_\Lambda$ (or, equivalently, through $\Omega_{\rm m} \simeq 1-\Omega_\Lambda$ in a flat Universe). This may hint at a dynamical behavior of dark energy.

In the Standard Model of particle physics, the  three generations of neutrinos are massless. However, neutrino detectors have shown that neutrinos undergo flavor oscillations \cite{Pontecorvo:1967fh,Davis:1968cp,Kajita:2016cak,LBNE:2013dhi}, which implies that they have a nonzero mass. 
However, oscillation experiments cannot measure the absolute neutrino mass scale. Instead, they determine squared mass differences between the three mass eigenstates, 
$\Delta m^2_{ij}=m_i^2 - m_j^2$.
This leads to two possible mass orderings, determined by the sign of \( \Delta m^2_{31} \). Normal hierarchy (NH)
corresponds to the mass pattern \( m_1 < m_2 \ll m_3 \), and inverted hierarchy (IH)
to \( m_3 \ll m_1 < m_2 \).
Recent global fits to neutrino oscillation data, as reported by \cite{Esteban:2020cvm,Esteban:2024eli,deSalas:2020pgw} at the \( 1\sigma \) confidence level, provide the following values for the mass-squared differences:
$\Delta m^2_{21}  = \left(7.49^{+0.19}_{-0.19}\right) \times 10^{-5}~\text{eV}^2$, and either
$\Delta m^2_{31}  = \left(2.513^{+0.020}_{-0.020}\right) \times 10^{-3}~\text{eV}^2 \quad \text{(NH)}$ or
$\Delta m^2_{32}  = \left(-2.484^{+0.020}_{-0.020}\right) \times 10^{-3}~\text{eV}^2 \quad \text{(IH)}$.
The minimum summed mass compatible with NH is around $\sum m_{\nu}\ge 0.058$eV, while IH requires $\sum m_{\nu}\ge 0.098$eV.

Neutrinos play a crucial role in governing the dynamics of the universe and cosmic structure formation. It is possible to determine the summed neutrino mass through various cosmological observations such as CMB, BAO and SNIa, through the effect of neutrinos on the background and perturbation evolution. The summed neutrino mass is one out of several cosmological parameters constrained by global fits to the data. As such, neutrino mass bounds from cosmology are intrinsically model dependent. For instance, there is a well-known correlation between the determination of the summed neutrino mass and that of the Hubble parameter or, possibly, of dynamical dark energy parameters~\cite{Sharma:2022ifr,Chudaykin:2022rnl,Reboucas:2024smm,Elbers:2024sha,Jiang:2024viw,Jiang:2024xnu,RoyChoudhury:2025dhe}. 
Therefore, it is generally expected that the current neutrino mass bounds might change in presence of new early or late time physics. 

Current constraints on the summed neutrino mass in the minimal scenario -- that is, in the seven-parameter $\Lambda$CDM+$\sum m_{\nu}$ model -- are strong and compatible with zero. In 2018, the combination of CMB data from {\it Planck}~\cite{Planck:2018vyg} and BAO data from several surveys like BOSS \cite{BOSS:2016wmc} led to $\sum m_\nu < 0.12~\text{eV}$ at the 95\% confidence level (CL) \cite{Planck:2018vyg,RoyChoudhury:2019hls}. This bound has been straighten by more recent CMB and BAO data.
The measurement of BAOs from the Dark Energy Spectroscopic Instrument (DESI) with Data Release 2 
(DESI DR2) \cite{DESI:2025ejh,DESI:2025zgx} returns a limit of 0.06 or 0.07~eV in combination with the {\it Planck} Data Release 4 (depending on which analysis pipeline and likelihood is used for {\it Planck}). This suggests the emergence of a possible new tension between laboratory and cosmological neutrino mass bounds -- as long as the $\Lambda$CDM model is assumed \cite{Jiang:2024viw}.

A well-known avenue for easing the Hubble tension is to assume that some new physics leads to a smaller value of the sound horizon at the time of recombination. This can be achieved mainly in two ways: by enhancing the expansion rate before and around recombination, for instance, with early dark energy (EDE)\cite{Karwal:2016vyq,Poulin:2018cxd,Poulin:2023lkg,Niedermann:2019olb,Chatrchyan:2024xjj,Garny:2025kqj,Sharma:2023kzr,SPT-3G:2025vyw}  or early modified gravity (EMG)\cite{Brax:2013fda,Braglia:2020auw,FrancoAbellan:2023gec}, or by changing the recombination history such that recombination takes place earlier\cite{2011arXiv1108.2517J,Jedamzik:2020krr,Schoneberg:2024ynd}. 

In cosmology, all tensions and parameter effects are inter-connected -- and in an optimistic interpretation, different tensions may turn out to be three different hints at the same alternative to $\Lambda$CDM. For instance, it has already been argued that models with a reduced sound horizon tend ease the tension on $\Omega_\Lambda$ and $\Omega_{\rm m}$ \cite{Chaussidon:2025npr,Mirpoorian:2025rfp}. Independently, it was shown that solving the $\Omega_\Lambda$ and $\Omega_{\rm m}$ tension by introducing dynamical (late) dark energy removes any neutrino mass tension \cite{DESI:2025ejh}. It is also legitimate to wonder whether reducing the sound horizon is a natural way to relax neutrino mass bounds: this is the main topic of this work.

Specifically, we will assume that the redshift of recombination is the same as in standard cosmology, while an unknown mechanism changes the early expansion rate and the value of the sound horizon at recombination. We will remain agnostic about this mechanism and its impact on early fluctuations in CMB photons. A similar approach was already introduced and used in \cite{Pogosian:2020ded,Lin:2021sfs,Wang:2025mqz,GarciaEscudero:2025lef}. References
\cite{Smith:2022iax,Farren:2021grl,Baxter:2020qlr} further extended this approach by including additionally some data on the full shape of the galaxy power spectrum marginalized over BAO features. We will analyze the generic impact of this category of models on the three possible tensions discussed above (Hubble, $\Omega_{\rm m}$ and neutrino mass).  

This article as organized as follows: In section \ref{sec:obs}, we summarize the impact of the summed neutrino mass on cosmological observables. In section \ref{sec:AgnosticEP}, we discuss how to use cosmological data while being agnostic on the sound horizon.
In section \ref{sec:Analysis}, we specify our observational data and methodology. In section \ref{sec:preliminary}, we carry a simplified analysis with a reduced number of free parameters in order to better understand the role of each data set. Our main results are presented in \ref{sec:results} and discussed in \ref{sec:discussion}.

\section{Impact of neutrino mass on cosmological observables}
\label{sec:obs}

\subsection{Impact on geometry}
Cosmic neutrinos decouple when the weak interaction rate falls below the Hubble expansion rate, at approximately $T_{\text{dec}} \approx 1\,\text{MeV}$. Being very light, neutrinos remain relativistic after decoupling, contributing to the radiation density in the early universe.  
They become non-relativistic during the matter-dominated era, around redshift $z_{\text{nr}} \approx 1890\, (m_\nu/\text{eV})$ 
\cite{Lesgourgues:2006nd,Hannestad:2006zg,Lesgourgues:2013sjj,Lattanzi:2017ubx,Loverde:2024nfi}.

Before recombination at $z=z_*$, neutrinos are still relativistic. Thus, their mass $\sum m_\nu$ has a negligible impact on the early expansion history and the comoving sound horizon:
\begin{equation}
r_{\rm s} = \int_{z_*}^{\infty} \frac{c_{\rm s}(z, \omega_{\rm b})}{H(z)} \, dz~,
\label{eq:rs}
\end{equation}
with $\omega_x\equiv \Omega_x h^2$
Instead, neutrino masses do impact the angular scale under which the sound horizon is observed in CMB data. 
The comoving angular diameter distance, called $D_M$ in DESI papers or $r_A$ in \class{} notations \cite{2011JCAP...07..034B}, is given by
\begin{equation}
    D_M(z)= f_k\left(\int_0^{z} \frac{c \, d\tilde{z}}{a_0 H(\tilde{z})}\right) \, .
    \label{eq:dm}
\end{equation}
In this work, we restrict ourselves to a spatially flat universe with $f_k(x)=x$.
The angular diameter distance to the surface of last scattering is given by $D_A(z_*)$ with:
\begin{equation}
D_A(z) = a \, D_M(z) = \frac{a_0}{1 + z} D_M(z).
\end{equation}
The Hubble parameter evolves as:
\begin{align}
\frac{H(z)}{H_0} = \Bigg[ 
& \Omega_{\rm cb}(1+z)^3 + \Omega_\gamma(1+z)^4 + \Omega_K(1+z)^2 \nonumber \\
& + \Omega_\nu \frac{\rho_\nu(z)}{\rho_{\nu,0}} + \Omega_{\rm DE} \frac{\rho_{\rm DE}(z)}{\rho_{\rm DE,0}} 
\Bigg]^{1/2}~,
\end{align}
featuring the respective contributions of CDM plus baryons, photons, curvature, neutrinos and dark energy.
In the relativistic regime ($z \gg z_{\text{nr}}$), the neutrino energy density scales as
\begin{equation*}
    \rho_\nu(z) = N_{\text{eff}} \cdot \frac{7\pi^2}{120} T_{\nu,0}^4 (1 + z)^4~,
\end{equation*}
where $T_{\nu,0}=(4/11)^{1/3} T_{\rm CMB}$ is the approximate neutrino temperature today in the instantaneous decoupling limit, and $N_{\text{eff}}$ is the effective neutrino number, given by 3.044 in the standard model \cite{Bennett:2020zkv,Froustey:2020mcq,Akita:2020szl}.
In the non-relativistic regime ($z \ll z_{\text{nr}}$), the neutrino energy density becomes\footnote{Strictly speaking, Eq.~(\ref{eq:NU_mass}) requires all neutrino species to be non-relativistic. It also holds approximately when one eigenstate has a negligible mass and the other two are non-relativistic.}\cite{Lesgourgues:2006nd,Hannestad:2006zg,Lesgourgues:2013sjj}
\begin{equation}
 \rho_\nu(z) = \left( \frac{\sum m_\nu}{93.14\, \text{eV} \cdot h^2} \right)  \frac{3H_0^2}{8\pi G} (1 + z)^3.   
 \label{eq:NU_mass}
\end{equation}
Plugging these relations into Eq.~(\ref{eq:dm}), one finds that for $z \ll z_{\rm nr}$ the angular distance is the same in all cosmologies with a given matter density $\omega_{\rm m}=\Omega_{\rm m} h^2 = \omega_{\rm cb} + \omega_\nu$ and DE density $\rho_{\rm DE}(z)$, independently of the summed neutrino mass $\sum m_\nu$ (or, equivalently, of the neutrino density fraction $f_\nu = \Omega_\nu/\Omega_{\rm m}$). However, for $z \gg z_{\rm nr}$, this is no longer the case, since the integrand in Eq.~(\ref{eq:dm}) runs over times at which neutrinos were ultra-relativistic. Thus, observables like the angular scale of the sound horizon in CMB maps depend on $\sum m_\nu$ (or $f_\nu$) on top of other parameters. This explains the well-known parameter correlation between $H_0$ and $\sum m_{\nu}$ in CMB data analyses\cite{Lesgourgues:2006nd,Lynch:2025ine,Loverde:2024nfi,Wang:2024hen,GarciaEscudero:2025lef}. 

According to the above discussion, the geometrical impact of neutrino masses cannot be probed by using only data from BAO or SNIa. These data are sensitive to $D_M(z)$, $D_A(z)$, and to the luminosity distance $D_L(z)=(1+z)^2 D_A(z)$ only up to redshifts $z \ll z_{\rm nr}$. As such, they can only return measurements of the total non-relativistic matter density $\omega_{\rm m}$ and of DE parameters (in the minimal model, of $\Omega_{\rm m} = 1 - \Omega_\Lambda$). 

The situation changes when CMB data is included, either in the form of anisotropy spectra $C_\ell$ or of a constraint on the angular scale of the sound horizon $\theta_{\rm s} = r_{\rm s}/D_M(z_*)$. Assuming a standard thermal history together with an accurate determination of $\omega_{\rm b}$ and of the redshift of equality, $r_{\rm s}$ can also be considered as fixed to about 145~Mpc. In that case, CMB data impose a given value of $D_M(z_*)$, that is, of the integral over $H^{-1}(z)$ from $z=0$ to $z=z_*$ that appears in Eq.~(\ref{eq:dm}). An increase in $\sum m_\nu$ needs to be compensated by a variation of at least one other cosmological parameter -- for instance, of $\Omega_{\rm m}$ -- in order to keep the integral fixed. Then, varying $\sum m_\nu$ affects indirectly the expansion history $H(z)$ also at low redshift $0<z<2$, as illustrated e.g. in Figure 1 of \cite{Pan:2015bgi}. This makes BAO and SNIa data indirectly sensitive to the geometrical effect of $\sum m_\nu$
\cite{Elbers:2024sha,Loverde:2024nfi,GarciaEscudero:2025lef,Lynch:2025ine}.

\subsection{Structure formation}

Even if neutrinos become non-relativistic in the matter-dominated era, they free stream with very high velocity and don't fall in small-scale gravitational potential wells created by CDM and baryons. On the other hand, they contribute to the expansion rate. As such, they shift the balance between the Hubble friction term and the gravity forces exerted on CDM and baryons, slowing down structure formation on scales smaller than the neutrino free-streaming scale \cite{Bond:1980ha,Hu:1997mj,Lesgourgues:2006nd,Lesgourgues:2013sjj,Loverde:2024nfi}. The impact of neutrino free streaming affects all observables related to the matter power spectrum, including the CMB lensing power spectrum.

\subsection{Early integrated Sachs-Wolfe effect}

Independently of the previous two effects, neutrinos affect the evolution of metric fluctuations during their transition form the relativistic to non-relativistic regime. For sufficiently large neutrino masses, this may happen just after photon decoupling and leave an imprint in the CMB temperature spectrum through the early integrated Sachs-Wolfe (ISW) effect \cite{Lesgourgues:2013sjj}. However, for masses in the range $m_\nu<0.3$~eV, this effect is very small and hard to detect. Additionally, in this work, we will not include information on the CMB temperature power spectrum shape, since our goal is to obtain constrains agnostic to early universe physics. Thus, the early ISW effect will not play a role in our study.

\section{Data combinations agnostic to early Universe Physics} 
\label{sec:AgnosticEP}

\subsection{Purely geometrical measurements}

If we want to remain agnostic to the mechanisms that define the sound horizon, we cannot use the full shape of CMB anisotropy spectra. The CMB temperature and polarization spectra are very sensitive to the physics of the early universe, that is, to the background, thermodynamical and perturbation evolution, both before and around recombination. The same mechanism leading potentially to a non-standard value of the sound horizon may impact the CMB spectra in a non-trivial way (some examples are provided by EDE models 
\cite{Karwal:2016vyq,Poulin:2018cxd,Niedermann:2019olb,Poulin:2023lkg,Chatrchyan:2024xjj,Garny:2025kqj} or stepped dark radiation models \cite{Aloni:2021eaq,Joseph:2022jsf,Schoneberg:2023rnx}). To remain agnostic about the sound horizon size, we thus need to work with compressed observables.

The observation of peaks in the CMB spectra can be compressed into a measurement of the angular scale of the sound horizon $\theta_{\rm s}$, given by the ratio of the comoving sound horizon sound at photon decoupling $r_{\rm s}$ over the comoving angular diameter distance at the same time, $D_M(z_*)$. It is sometimes believed that the measurement of $\theta_{\rm s}$ is a purely geometrical measurement, directly given by the observed scale of the peak in the CMB power spectra, and thus model-independent. This is oversimplified. For instance, the relation between the scale of the first acoustic peak in the temperature spectrum, $\ell_{\rm p}$, and $\theta_{\rm s}$ is far more complicated than $\theta_{\rm s}=\pi/\ell_{\rm p}$. The true relation depends on the amplitude of the ``neutrino drag'' effect \cite{Bashinsky:2003tk,Lesgourgues:2013sjj,Baumann:2019keh}, on the early ISW (affecting mainly the first CMB temperature peak scale), on the ratio between the Sachs-Wolfe and Doppler contributions to the total anisotropy spectrum, and potentially on new physics not considered here. 
 
 This model dependence can be mitigated by inferring $\theta_{\rm s}$ from a global fit to all observed temperature and polarization peaks, when not only the six $\Lambda$CDM parameters but also $N_{\rm eff}$ are marginalized over. In the case of {\it Planck} data, such a fit provides the observational constraint\footnote{To extract this number, we used {\sc MontePython} \cite{Audren:2012wb,Brinckmann:2018cvx} and \class{} \cite{2011JCAP...07..034B} to fit the {\it Planck} baseline TT,TE,EE likelihoods (without lensing), assuming a 7-parameter $\Lambda$CDM+$N_{\rm eff}$ cosmology and one massive neutrino with $m_\nu=0.06$~eV.} 
    \begin{equation}
    100 \, \theta_{\rm s} =\frac{100 \, r_{\rm s}}{D_M(z_*)}
    =1.04223\pm 0.00055~ (68\%{\rm CL}),
    \label{eq:thetasplanck}
    \end{equation}
which can be considered as very weakly model dependent. Thus, this measurement can be used in a sound-horizon agnostic approach as long as we consider $r_{\mathrm{s}}$ as a free parameter and use Eq.~\eqref{eq:thetasplanck} to constrain $D_M(z_*)$. If we wanted to be additionally agnostic to the physics of recombination, we would also treat $z_*$ as a free parameter, but in this work we assume a standard recombination history and let \class{} compute $z_*$ in each cosmology like in the standard model (with a small dependence of $z_*$ on $\omega_{\rm b}$).  

Similarly, with galaxy redshift surveys, the reconstruction of the galaxy matter power spectrum or two-point correlation function in directions orthogonal to the line-of-sight and in a redshift bin centered on a given $z_i$ provides an estimate of 
\begin{align}
    \theta_{\rm s}(z_i)= \frac{r_{\rm d}}{D_M(z_i)}~,
    \label{eq:thetas_bao}
\end{align}
where 
\begin{equation}
r_{\rm d} = \int_{z_{\rm d}}^{\infty} \frac{c_{\rm s}(z, \omega_{\rm b})}{H(z)} \, dz~,
\end{equation}
and $z_{\rm d}$ is the redshift of baryon drag. The same discussion holds as in the case of the CMB: the measurement of the angles $\theta_{\rm s}(z_i)$ is not truly direct nor model-independent. What the data directly measures are features in the galaxy correlation function or power spectrum. Inferring $\theta_{\rm s}(z_i)$ from the power spectrum requires a proper estimate of the phase of the acoustic oscillations imprinted in the matter power spectrum~\cite{Baumann:2019keh,Chen:2024tfp}, which depends on an accurate modeling of various effects such as neutrino drag. Inferring $\theta_{\rm s}(z_i)$ from either the correlation function or power spectrum also requires an accurate modeling of non-linear gravitational clustering effects \cite{Chen:2024tfp}. Observational collaborations like DESI \cite{DESI:2025qqy} take these effects into account and provide an estimate of $\theta_{\rm s}(z_i)$ valid at least for a $\Lambda$CDM cosmology, but applicable in practice to more extended models. The reason is that  the dependence of the non-trivial effects mentioned above on extended cosmological parameters is typically negligible compared to current observational errors on $\theta_{\rm s}(z_i)$, as discussed for instance in \cite{Thepsuriya:2014zda} and Appendix A of \cite{Schoneberg:2019wmt}. Thus, in our agnostic approach, we can use the DESI measurements of the BAO angular scale, as long as we consider $r_{\rm d}$ as a free parameter and use each data point to constrain $D_M(z_i)$.

The same discussion applies to the measurement of the BAO scale along the line of sight, which returns a nearly model-independent estimate of the ratio
\begin{align}
\Delta z(z_i) \equiv \frac{r_{\rm d}}{D_H(z_i)} = \frac{H(z_i) r_{\rm d}}{c}~.
\end{align}  
In our agnostic analysis, we will treat $r_{\rm d}$ as a free parameter and use each DESI measurement of $\Delta z(z_i)$ to constrain $H(z_i)$.

\subsection{From geometrical measurements to an uncalibrated inverse distance ladder likelihood
\label{sec:agnostic}}

There are two ways to implement a likelihood that incorporates the geometrical measurements discussed in the previous section while remaining agnostic about the value of the sound horizon. In both cases, such a likelihood provides a measurement of the distances $D_A(z)$ and $D_H(z)$ at $z \leq z_*$, calibrated to a standard ruler at high redshift -- the sound horizon. This is the principle of the inverse distance ladder (IDL) approach, to be contrasted with the distance ladder method, for which distances are calibrated to standardized objects at very low redshift. Remaining agnostic about the calibrator, in this case the sound horizon, amounts in building an {\it uncalibrated} IDL likelihood -- in the same way as a distant supernovae likelihood marginalized over the standardized SNIa magnitude $M_{\rm B}$ is an {\it uncalibrated} distance ladder (DL) likelihood.\\

{\bf Method I.} Firstly, one can  combine the previous measurements to eliminate the dependence on $r_{\rm s}$ and $r_{\rm d}$ up to a large extent. We define the following ratios of CMB and galaxy data:
\begin{align}
      R_M(z_i) &\equiv 
      100\, \frac{r_{\rm s}}{r_{\rm d}} \frac{D_M(z_i)}{D_M(z_*)}~,
      \label{eq:RM}\\
      R_H(z_i) &\equiv 
      \frac{D_M(z_i)}{D_H(z_i)}~,
      \label{eq:RH}
\end{align}
and we fit these ratios to the observed values of, respectively,
\begin{equation} 
      \frac{100\, \theta_{\rm s}}{\theta_{\rm s}(z_i)}~, \qquad
      \frac{\Delta z (z_i)}{\theta_{\rm s}(z_i)}~,
      \label{eq:RH}
\end{equation}
assuming Gaussian errors. In some special cases, the observational collaboration only provides an estimate of the spherically-averaged BAO scale. In that case, we define
\begin{equation}
        R_V(z_i) \equiv 100 \, \frac{r_{\rm s}}{r_{\rm d}} \frac{D_V(z_i)}{D_M(z_*)}\, ,
        \end{equation}
where $D_V(z) = \left[ z \, D_H(z) \, D_M^2(z) \right]^{1/3}$, and we fit this ratio to the observed value of 
\begin{equation}
        100 \, \theta_{\rm s} \frac{D_V(z_i)}{[z_i \, \Delta z(z_i)\, \theta_{\rm s}^2(z_i)]^{1/3}}~.
        \end{equation}
All these ratios directly probe the geometry and expansion rate of the Universe at $z \leq z_*$. 

More precisely, $R_M(z_i)$ and $R_V(z_i)$ still depends on the factor\footnote{For an accurate calculation of ${r_{\rm s}}/{r_{\rm d}}$, it is necessary to define precisely the redshifts $z_*$ and $z_{\rm d}$ that enter into the calculation of the sound horizon at the recombination or baryon drag time. Here we define $z_*$ as the redshift at the maximum of the photon visibility function, $g'(z_*)=0$, and $z_{\rm d}$ as the redshift at which the baryon optical depth $\kappa_{\rm b}(z)$ crosses one. The baryon optical depth is related to that of photons, $\kappa$, through $\kappa_{\rm b}=\kappa/R$, with $R=3 \rho_{\rm b}/(4 \rho_\gamma)$.} ${r_{\rm s}}/{r_{\rm d}}$, but even in extended cosmological models, this ratio remains very close to 0.983 and depends very weakly on cosmological parameters -- such that this dependence can be neglected compared to BAO measurement errors. 

For instance, when fitting the $\Lambda$CDM model to a combination of CMB, BAO and SNIa data (from {\it Planck} 2018, DESI BAO DR2 and and Pantheon+), one gets $r_{\rm s}/r_{\rm d} =
0.98283_{-0.00027}^{+0.00025}$ at the 68\% CL. The impact of this uncertainty on $R_M(z_i)$ or $R_V(z_i)$ remains at least two orders of magnitude smaller than the current observational error on these ratios (in any redshift bin $i$). One may fear that in some extended cosmologies, the dependence of $r_{\rm s}/r_{\rm d}$ over cosmological parameters could be enhanced. For this, one would need to consider very specific models in which photons or baryons experience non-standard interactions or non-thermal distortions that shift their respective decoupling times by a different amount. Models that just shift the overall recombination history, for instance due to small-scale primordial magnetic fields \cite{2011arXiv1108.2517J,Jedamzik:2020krr} or shifts in fundamental constants \cite{Sekiguchi:2020teg}, change both $z_*$ and $z_{\rm d}$ in such a way that $r_{\rm s}/r_{\rm d}$ is not significantly affected. Other models that change the early expansion rate, such as EDE or EMG, do not affect $z_*$ and $z_{\rm d}$, but shift both $r_{\rm s}$ and $r_{\rm d}$ in such a way that their ratio is again nearly the same. For instance, when fitting the EDE model of Ref.~\cite{Niedermann:2019olb} to 
the same combination of CMB, BAO and Supernoave data, one gets $r_{\rm s}/r_{\rm d} =0.98334_{-0.00051}^{+0.00042}$ at the 68\% CL. This is compatible with the ratio found in the $\Lambda$CDM case, and the impact of this standard deviation is still at least 30 times smaller than the current observational error on any $R_M(z_i)$ or $R_V(z_i)$.

Thus, in an MCMC run, when computing $R_M(z_i)$ or $R_V(z_i)$ in a given cosmology, one can substitute $r_{\rm s}/r_{\rm d}$ with either a fixed factor 0.983 or the value computed by \class{} in the same cosmology (but assuming standard early universe physics). Neither choice is expected to bias the result\footnote{We checked explicitly with dedicated runs that these two options return the same results, and we adopt the latter method by default.} nor introduce a model dependence. We conclude from this discussion that constraints on $R_M(z_i)$ or $R_V(z_i)$ can effectively be seen as pure measurements of distance ratios.\\

{\bf Method II.} Instead of constructing ratios to eliminate $r_{\rm s}$ and $r_{\rm d}$, we can simply treat these quantities as free parameters and marginalize over them with a given prior. In that case, one may use the standard DESI BAO likelihood and a Gaussian likelihood for $\theta_{\rm s}$, just substituting the value of $r_{\rm s}$ or $r_{\rm d}$ computed by \class{} (under the assumption of a standard cosmology) with free values.

It would in principle be possible to marginalize over $r_{\rm s}$ and $r_{\rm d}$ independently. However, the discussion of the previous paragraphs shows that it is reasonable to assume that, even in extended cosmologies, the prediction ${r_{\rm s}}/{r_{\rm d}}\simeq 0.983$ holds up to some negligible uncertainty. Thus, it is possible to fluctuate only one of these two parameters and infer the other, using either a fixed factor $r_{\rm s}/r_{\rm d} = 0.983$ or the value computed by \class{} assuming a standard cosmology. As discussed before, these two options make no difference in practice. In our implementation, we fluctuate $r_{\rm s}$ with a flat prior (within a prior range wide enough not to influence the results) and we use the value of $r_{\rm s}/r_{\rm d}$ computed by \class{} to relate $r_{\rm d}$ to $r_{\rm s}$.

This Method II was recently adopted in \cite{GarciaEscudero:2025lef}. The two methods are expected to be equivalent up to prior volume effects. However, the data are constraining enough to limit such effects. To check this, we performed some runs using Methods I or II, and found nearly identical results (see Appendix \ref{app:methods}). In what follows, we adopt Method II as our baseline (this choice is arbitrary).

\subsection{Other observational constraints agnostic to the sound horizon}

To complete the information on cosmological distances coming from IDL measurements, we can use luminosity distance measurements from type Ia supernova (SNIa), probing $D_{L}(z)=(1+z)D_{M}(z)=(1+z)^2 D_A(z)$ up to an unknown calibration parameter, the standardized magnitude $M_B$. We will not use any prior on $M_B$ inferred from DL techniques, but we will examine whether the values of $M_B$ returned by our analysis are compatible with the determination of this parameter by DL experiments such as SH0ES.

Big Bang Nucleosynthesis (BBN) is relatively insensitive to the details of the cosmological model apart from the expansion rate at $z\sim {\cal O}(10^6)$, that can be expressed in most cases as a function of the effective neutrino number at BBN, $N_{\rm eff}^{\rm BBN}$, and the baryon-to-photon density ratio, parametrized through $\omega_{\rm b}$. Here, we assume that the new physical mechanism that may alter standard predictions for the sound horizon does not affect BBN. Thus, we can use a Gaussian BBN likelihood for $\omega_{\rm b}$, inferred from the primordial abundance of Deuterium and Helium-4 with $N_{\rm eff}^{\rm BBN}$ fixed to 3.044.\footnote{Note that, in principle, we could be more conservative, and assume a prior on $\omega_{\rm b}$ inferred after marginalization over an unknown $N_{\rm eff}^{\rm BBN}$ (constrained solely by primordial element abundances). In practice, this would increase the uncertainty on $\omega_{\rm b}$ by about 10\% and leave our main results unaffected.}

We already argued that we cannot use data from CMB anisotropy spectra on sub-degree scales, because of their sensitivity to the details of the cosmological model before and around recombination. However, this is not the case for anisotropies on super-degree scales. Under very general assumptions, it is reasonable to assume that the large-scale branches of the CMB temperature and polarization spectra provide information on the primordial power spectrum of adiabatic perturbations, corrected by reionization effects and, to a lesser extent, by the late ISW effect. Large-scale CMB anisotropies thus provide information on $A_{\rm s}$, $n_{\rm s}$, $\tau_{\rm reio}$, $\Omega_{\Lambda}$, and are actually sufficient to constrain independently the first three parameters. This means that even in an approach agnostic to the sound horizon, it is meaningful to use a prior on $A_{\rm s}$ and $n_{\rm s}$ derived from CMB observations after marginalization over  $\tau_{\rm reio}$ and $\Omega_{\Lambda}$. 

Finally, if we assume that the physical mechanism leading to a non-standard sound horizon is limited to early times, we expect that the formation of dark matter structures on large scales proceeds exactly like in standard cosmology. Currently, the best probes of this formation history come from galaxy weak lensing (based on cosmic shear surveys such as DES \cite{DES:2021wwk,DESCollaboration:2025udj} or KiDS \cite{Wright:2025xka,Harnois-Deraps:2024ucb}) and CMB lensing extraction (based on data from {\it Planck} \cite{Carron:2022eyg}, ACT \cite{ACT:2023dou,ACT:2023kun} or SPT \cite{SPT-3G:2024atg}). The former can be summarized as a prior on the weak lensing amplitude parameter $S_8=\sigma_8\sqrt{{\Omega_{\rm m}}/{0.3}}$, which is straightforward to implement in our analysis. The latter provides constraints on the lensing potential power spectrum, $C_\ell^{\phi\phi}$, which is sensitive to the broadband shape of the matter power spectrum but not to the sound horizon, since BAO features get washed out by projection effect \cite{BOSS:2016hvq}.

The use of a CMB lensing likelihood in our context requires a few words of caution. The measurement of $C_\ell^{\phi\phi}$ is inferred from a weighted sum of 4-point correlation functions of temperature and polarization maps. Ideally, we would like to consider this information as independent of the temperature and polarization spectra (that is, 2-point correlation functions), and only dependent on the growth of structure during matter and dark energy domination. As such, it would be perfectly suited for our agnostic approach. However, lensing extraction is performed under  the assumption of some fiducial temperature/polarization spectra. In our agnostic approach, we don't have full theoretical predictions for the temperature/polarization spectra of the unspecified models that incorporate new physics in the early universe and change the sound horizon. The best we can do is to assume that the measurements of the temperature and polarization spectra by current experiments are correct, such that any model allowed by the data has some $C_l^{TT}$, $C_l^{TE}$, $C_l^{EE}$ not too far from the {\it Planck} best-fit $\Lambda$CDM model. Then, it is legitimate to perform lensing extraction using the {\it Planck} best-fit spectra as fiducial spectra, and to propagate the uncertainty on the observed temperature/polarization spectra as an additional uncertainty on the reconstructed lensing spectrum $C_\ell^{\phi\phi}$. This is precisely what some versions of CMB likelihoods do (for {\it Planck}, the CMB-marginalized lensing likelihood; for ACT+{\it Planck}, the likelihood associated to the option \texttt{lens\_only=True}). Here, we will stick to such versions.\footnote{We are grateful to the authors of \cite{GarciaEscudero:2025lef} for stressing the importance of using such versions: with baseline CMB likelihoods, the results would be biased by unphysical corrections to $C_\ell^{\phi\phi}$ and would return significantly different results -- in particular, much lower values of $H_0$.}

In this work, to stay on the more conservative side, we do not include information on the full shape of the galaxy power spectrum. References
\cite{Smith:2022iax,Farren:2021grl,Baxter:2020qlr} suggest a very interesting way to include such information while marginalizing the galaxy spectrum over BAO features. While this method removes any direct sensitivity to the sound horizon, it still assumes that the broadband galaxy power spectrum can accurately be predicted under the same assumptions as in standard cosmology. Since the new physics changing the sound horizon may also change the background and perturbation evolution around the time of radiation-to-matter equality, discarding information from the galaxy power spectrum is way to be even more agnostic about this new physics.

\section{Analysis details}
\label{sec:Analysis}

\subsection{Data}
\label{sec:data}

We use the following data sets in our analysis.
\begin{itemize}
    \item {\bf calibrated IDL:}
    This IDL likelihood is built from two pieces accounting for CMB peak and BAO observations. The first one is a Gaussian likelihood for the measurement of the angular scale of the sound horizon at recombination, $\theta_{\rm s}$,  extracted from {\it Planck} data fitted with a $\Lambda$CDM+$N_{\rm eff}$ model, see Eq.~(\ref{eq:thetasplanck}).
    The second one is the likelihood that accounts for the BAO data from DESI Data Release 2 (DR2), based on different tracers in the redshift range $0.1<z<4.2$ \cite{DESI:2025zgx}:
BGS ($z_0 = 0.295$);
LRG1 ($z_1 = 0.510$);
LRG2 ($z_2 = 0.706$);
LRG3+ELG1 ($z_3 = 0.934$);
ELG2 ($z_4 = 1.321$);
QSO ($z_5 = 1.484$);
Lyman-$\alpha$ ($z_6 = 2.330$).
For the BGS tracer, only the ratio $D_V/r_{\rm d}$ is measured. For other tracers, the likelihood incorporates measurements of $D_M/r_{\rm d}$ and $D_H/r_{\rm d}$ (see Table~IV of \cite{DESI:2025zgx}). The first likelihood relies on a theoretical prediction for $r_{\rm s}$ and the second one for $r_{\rm d}$; these distances are computed by \class{} for each model according to standard physical assumptions.
    
    \item {\bf uncalibrated IDL:} To get an IDL likelihood agnostic to the sound horizon, we follow be default the Method II described in section~\ref{sec:agnostic}. We use the same CMB and BAO likelihoods as in the calibrated IDL case, but with $r_{\rm s}$ treated as a free nuisance parameter with a flat prior, and $r_{\rm d}$ inferred from $r_{\rm s}$. For comparison, we also implemented Method I, introducing a Gaussian likelihoods for each of the ratios defined in section~\ref{sec:agnostic}. Table~\ref{tab:cosmo_obs} summarizes the observed value and standard error for each ratio. We infer such measurements from Eq.~(\ref{eq:thetasplanck}) and Table~IV of \cite{DESI:2025zgx}, with error bars assuming products of Gaussian distributions, ${A}{B}\pm {A}{B}\sqrt{\frac{\sigma_{A}^2}{A^2}+\frac{\sigma_{B}^2}{B^2}}$. Appendix \ref{app:methods} shows that the two methods provide very similar results.
    \begin{table}[ht]
\centering
\begin{tabular}{|l|c|c|c|c|}
\hline
\textbf{Type} & $z_i$ & 
\textbf{$R_M(z_i)$}
& 
\textbf{$R_V(z_i)$} 
& 
\textbf{$R_H(z_i)$} \\
\hline
BGS     & 0.295   & $-$ & 
$8.28\pm0.08$
& $-$ \\
LRG1    & 0.510   &   $14.16\pm 0.17$ & $-$
& $0.622\pm 0.017$ \\
LRG 2    & 0.706   &  $18.08\pm 0.18$&
$-$
& $0.892\pm 0.021$ \\
LRG3+ELG1& 0.934   & $22.49\pm 0.16$&
$-$
& $1.223\pm 0.019$ \\
ELG2 & 1.321     & $28.76\pm 0.33$&
$-$
& $1.948\pm 0.045$ \\
QSO & 1.484    & $31.80\pm 0.79$&
$-$
& $2.386\pm 0.136$ \\
Lyman-$\alpha$  & 2.330     & $40.63\pm 0.55$&
$-$
&  $4.518\pm 0.097$ \\
\hline
\end{tabular}
\caption{Measurement of distance ratios entering in our agnostic IDL likelihood when following Method I. The definition of the ratios $R_M$, $R_H$ and $R_V$ is given in Sec.~\ref{sec:agnostic} and the values are inferred from DESI DR2 data. Each measurement is treated as Gaussian and the error bars stand for standard deviations.}
\label{tab:cosmo_obs}
\end{table}

    \item {\bf uncalibrated SNIa:} We use the luminosity distance measurements of uncalibrated SNIa, treating the standardized magnitude $M_B$ as a nuisance parameter that gets marginalized over.
    As a baseline, we use the Pantheon$+$ sample \cite{Brout:2022vxf}. For comparison, we perform some of our runs with the DES Year-5 SNIa dataset~\cite{DES:2024jxu}.

    \item {\bf BBN:} We use a measurement of the baryon density inferred from up-to-date measurements of the primordial abundance of Deuterium and Helium-4 assuming standard BBN,
    $\omega_{\rm b}=0.02218\pm0.00055$
    (68\% CL) \cite{Schoneberg:2024ifp} (see also \cite{Pisanti:2020efz}).

    \item {\bf Cosmic shear:} We use a split-normal prior on 
    $S_8 = 0.815^{+0.016}_{-0.021}\ (\text{68\% CL})$
 inferred from the KiDS Legacy survey \cite{Wright:2025xka}. In some places, we show for comparison the results obtained when switching to the prior $S_8=0.776\pm0.017$ (68\%CL) inferred from the analysis of DES Year-3 \cite{DES:2021wwk}.

    \item {\bf $A_{\rm s}$, $n_{\rm s}$ priors:} In some of our runs, we use some Gaussian priors on $\ln A_{\rm s}$ and/or $n_{\rm s}$ inferred from {\it Planck} 2018 data after marginalizing over the other five parameters of a $\Lambda$CDM+$N_{\rm eff}$ model,
    $\ln 10^{10} A_{\rm s}=3.0372\pm0.0194$ (68\%CL) and 
    $n_{\rm s}=0.9587\pm0.0091$ (68\%CL).
    
    \item {\bf CMB lensing:} We use the measurement of the lensing power spectrum $C_{l}^{\phi\phi}$ inferred from a combination of ACT DR6 and  {\it Planck} NPIPE data \cite{ACT:2023dou,ACT:2023kun,Carron:2022eyg}. As mentioned before, we set the flag \texttt{lens\_only} to \texttt{True} for consistency.   

   \item {\bf direct DL}: At some point in the final discussion section \ref{sec:discussion}, we will combine the previous data with the SH0ES measurement of the standardized SNIa likelihood, implemented as a gaussian prior \cite{Riess:2021jrx}
   \begin{equation}
     M_{B}=-19.253\pm 0.027 \quad (68\%\mathrm{CL})~.
     \label{eq:SHOES}
   \end{equation}

   \item {\bf $\beta$-decay}:  The KATRIN experiment is sensitive to the $\beta$-decay mass $m_\beta$. The collaboration currently  obtains an upper bound $m_\beta<0.45$~eV (90\%CL) \cite{KATRIN:2024cdt}. For a mass close to this threshold, neutrinos must obey to the degenerate mass scheme for which $\sum m_\nu \simeq 3 m_\beta$. At the end of the discussion section \ref{sec:discussion}, we will use a prior on the summed neutrino mass inspired from KATRIN, taking the form of a half-Gaussian peaking in zero and with standard deviation $\sigma(\sum m_\nu)=0.85$~eV.\footnote{Identifying the KATRIN bound on $m_\beta$ to be a 1.6$\sigma$ bound, we divide 0.45~eV by 1.6 and multiply by the number of neutrinos to get $\sigma(\sum m_\nu)=0.85$~eV. Since this deviation is much larger than the smallest value of $\sum m_\nu$ compatible with oscillation data, $\sum m_\nu\simeq 0.06$~eV,  assuming that the prior peaks in zero or near $0.06$~eV would make no difference in practice.} 
    
\end{itemize}

\subsection{Parameter inference}

We use the publicly available Einstein-Boltzmann solver \class{}\footnote{\url{https://class-code.net}\\ \url{https://github.com/lesgourg/class_public}}\cite{2011JCAP...07..034B,2011arXiv1104.2932L} interfaced with the MCMC sampler {\sc MontePython}\footnote{\url{https://github.com/brinckmann/montepython_public}} \cite{Audren:2012wb,Brinckmann:2018cvx}. We sample the parameter space using the Metropolis Hastings algorithm. For each MCMC, we run 6 parallel chains and enforce a Gelman-Rubin criterium $\max[R-1]<0.03$ for good convergence  \cite{Gelman:1992zz}.

Our runs assume either a cosmological constant or dynamical dark energy with the effective Chevallier-Polarski-Linder (CPL) parametrization \cite{Chevallier:2000qy,Linder:2002et}. In the $\Lambda$CDM case, we assume flat priors on the cosmological parameters ($\omega_{\rm b},\omega_{\rm cdm}, H_0,\ln 10^{10} A_{\rm s},n_{\rm s},\sum m_{\nu}$). The $w_0 w_a$CDM case features two additional parameters $(w_0,w_a)$. Our prior ranges are wide enough not to be reached by any 95\% confidence contour (uninformative prior), with the exception of $\sum m_{\nu}>0$ -- see \ref{tab:priors} for full list of priors used in analysis. 
None of our likelihoods is sensitive to the optical depth to reionization. We assume three degenerate massive neutrinos and ensure that before the non-relativistic transition the radiation density matches $\rm N_{\rm eff}=3.044$.

\begin{table}[h!]
\centering
\begin{tabular}{lc}
\hline
\textbf{Parameter} & \textbf{Prior Range} \\
\hline
$H_0$ [km/s/Mpc] & [50, 90] \\
$\omega_{\rm b}$ & [0.017, 0.027] \\
$\omega_{\rm cdm}$ & [0.08, 0.18] \\
$\ln(10^{10} A_{\rm s})$ & [2.4, 3.4] \\
$n_{\rm s}$ & [0.90, 1.1] \\
$\sum m_\nu/3$ [eV] & [0, 10] \\
$w_0$ & [-3, 2] \\
$w_a$ & [-3, 2] \\
$M_B$ & [-20, -18] \\
$r_{\rm s}$ [Mpc] & [100, 200] \\

\hline
\end{tabular}
\caption{Priors on cosmological parameters.}
\label{tab:priors}
\end{table}

\section{Preliminary analysis}
\label{sec:preliminary}

To understand the role and the constraining power of each data set, we run a preliminary analysis in which we stick to the sound-horizon-agnostic version of the minimal $\Lambda$CDM model (with $\sum m_\nu=0.06$~eV, $w_0=-1$, $w_a=0$) while fixing the primordial spectrum parameters to $\ln (10^{10} A_{\rm s})=3.0372$, $n_{\rm s}=0.9587$, and the baryon density to $\omega_{\rm b}=0.02218$. We thus vary only three free cosmological parameter with top-hat priors: $(r_{\rm s}, H_0, \omega_{\rm cdm})$. In this parameter space, $\Omega_{\rm m}=(\omega_{\rm cdm}+ \omega_{\rm b})/h^2$ appears as a derived parameter. 

\begin{figure}
    \centering
    \includegraphics[width=0.99\linewidth]{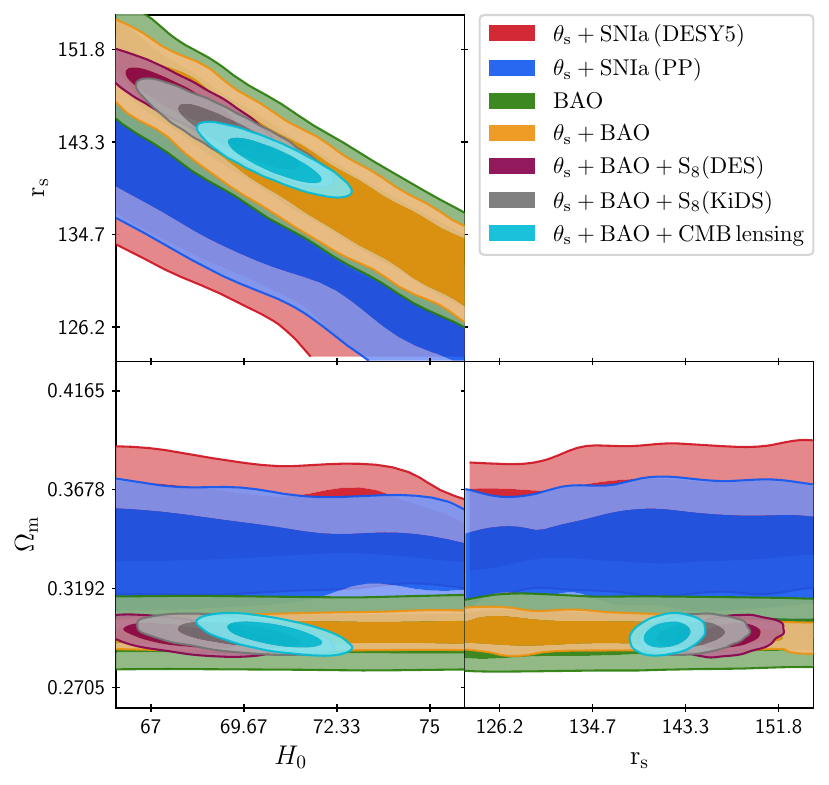}
    \caption{Assuming a $\Lambda$CDM model with a free sound horizon and fixed values of $(\omega_{\rm b}, A_{\rm s}, n_{\rm s})$, 68\% confidence contours for ($r_{\rm s}$, $\Omega_{\rm m}$, $H_0$) with different data sets (see legend and text).}
    \label{fig:preliminary}
\end{figure}

We show in Figure \ref{fig:preliminary} the 2D confidence contours on $(r_{\rm s}, \Omega_{\rm m}, H_0)$ using various combinations of uncalibrated distance data (uncalibrated BAO, uncalibrated $\theta_{\rm s}$, uncalibrated SNIa) and probes of the growth of structures ($S_8$ from cosmic shear, CMB lensing).\\

{\bf Constraints on $\Omega_{\rm m}$.} As expected, when only uncalibrated distance data are used, there is no way in which the Hubble rate and the sound horizon could be determined: they are fully degenerate with each other and anti-correlated (a reduced sound horizon yields a larger Hubble parameter). On the other hand, uncalibrated distance data are sensitive to the dark energy parameters that impact the shape of the expansion rate $H(z)$ at low redshift (but not its overall normalization given by $H_0$), and thus, also the shape of $d_A(z)$ and $d_L(z)$. In the present case, the only parameter of this type is $\Omega_\Lambda=1-\Omega_{\rm m}$. We can check in Figure \ref{fig:preliminary} that the uncalibrated distance data provide a constraint on this parameter (and only this one). Regarding the constraint on $\Omega_{\rm m}$, we observe that uncalibrated BAO and $\theta_{\rm s}$ data are very consistent with each other, since the BAO+$\theta_{\rm s}$ contours lies in the middle of the BAO-alone contours. We also confirm that there is a mild (slightly above 2$\sigma$) tension between the confidence intervals for $\Omega_{\rm m}$ derived from either 
``BAO (DESI) + $\theta_{\rm s}$'' or ``SNIa (DES-Y5) + $\theta_{\rm s}$'': this is the well-known $\Omega_{\rm m}$ tension that has emerged recently and that might be interpreted as a hint for dynamical dark energy. The combination ``SNIa (Pantheon$+$) + $\theta_{\rm s}$'' also prefers smaller $\Omega_{\rm m}$ values than ``BAO (DESI) + $\theta_{\rm s}$'', but in this case there is still an agreement at better than 2$\sigma$. In what follows, we will use the Pantheon$+$ SNIa data in our baseline analysis.\\

{\bf Constraints on $(H_0, r_{\rm s})$.} To get bounds on the other two free parameters $(H_0, r_{\rm s})$, we must involve data on the growth of structure, which provide information on the duration of matter domination and the age of the Universe, and thus, on $H_0$ (once $\Omega_{\rm m}$ has been fixed by distance data). Due to the correlation between $H_0$ and $r_{\rm s}$ imposed by distance data, structure growth data will also provide bounds on the sound horizon. There are two ways to get such additional information:
\begin{itemize}
\item The first one is to compare the amplitude of primordial fluctuations set by $A_{\rm s}$ to the amplitude of current matter fluctuations set by $S_8$. Since these parameters refer to the amplitude of the spectrum of matter fluctuations at different pivot scales, we also need some information on $n_{\rm s}$. Finally, baryons and massive neutrinos could reduce the growth of structures on the scales probed by the $S_8$ prior, so $\omega_{\rm b}$ and $\sum m_\nu$ are important too. In our preliminary analysis, we fix $A_{\rm s}$, $n_{\rm s}$, $\omega_{\rm b}$ and $\sum m_\nu$, such that a prior on $S_8$ is sufficient to get bounds on $H_0$.
\item The second way is to compare the amplitude of primordial fluctuations set by $A_{\rm s}$ to the amplitude of the CMB lensing potential. Compared to an $S_8$ prior, CMB lensing data contain additional information on the shape of the metric fluctuation or matter power spectrum, and thus, on $\omega_{\rm b}$, $n_{\rm s}$ and $\sum m_\nu$. Here, since we are fixing $A_{\rm s}$, $n_{\rm s}$, $\omega_{\rm b}$ and $\sum m_\nu$, the CMB lensing likelihood can be used to constrain $H_0$.
\end{itemize}

Figure \ref{fig:preliminary} shows how $H_0$ and $r_{\rm s}$ get bounded when we add to uncalibrated IDL data (uncalibrated BAO and $\theta_{\rm s}$) either an $S_8$ prior inferred from DES or KiDS, or CMB lensing data. In each case, we get a relatively narrow range of possible $H_0$ and $r_{\rm s}$ values. The $H_0$ and $r_{\rm s}$ posteriors from these runs do not provide reliable predictions, since they are not marginalized over important unknown parameters such as $(A_{\rm s}, n_{\rm s}, \omega_{\rm b})$ and potentially $\sum m_\nu$. It is still interesting to see that the $S_8$ measurement from DES is pushing for a lower range of $H_0$ values, while CMB lensing data is pushing for a higher one and the $S_8$ measurement from KiDS stands in the middle. However, these contours are all compatible with each other at the 2$\sigma$ level. 
The difference between the $H_0$ posteriors from DES and KiDS is consistent with the fact that DES prefers a slightly lower range for $S_8$. If $H_0$ is lower, the universe is younger, matter domination is shorter, and perturbations experience less growth.
In what follows, we will adopt the more recent KiDS result as our baseline for the $S_8$ prior, while keeping in mind that the DES result would prefer slightly lower $H_0$ values.

Figure \ref{fig:preliminary} suggests that the $S_8$ prior and CMB lensing data have roughly the same constraining power. This may no longer be the case once we marginalize the results over $n_{\rm s}$, $\omega_{\rm b}$ and $\sum m_\nu$. These parameters can be constrained by the shape information contained in CMB lensing data, but they are very degenerate with $H_0$ when using only the $S_8$ prior. We therefore expect that in the full analysis, CMB lensing becomes more constraining than the $S_8$ prior. This also mitigates the impact of choosing the KiDS rather than DES prior on $S_8$ as our baseline.

\section{Results}
\label{sec:results}

We define the following data combinations:
\begin{itemize}
    \item {\bf Agnostic baseline:} This stands for the combination of uncalibrated IDL data (based on BAO from DESI Y3 and $\theta_{\rm s}$ from {\it Planck}), uncalibrated SNIa data (based on Pantheon$+$), BBN (implemented as an $\omega_{\rm b}$ prior), and cosmic shear (implemented as an $S_8$ prior based on KiDs Legacy).

    \item {\bf Standard baseline:} This is the same combination of likelihoods as the agnostic baseline, but with the uncalibrated IDL likelihood replaced by the calibrated one, that is, using the standard predictions for $r_{\rm s}$ as a function of other cosmological parameters.
\end{itemize}

For our main results, we fit each of the two models $\Lambda$CDM+$\sum m_\nu$ and $w_0 w_a$CDM+$\sum m_\nu$ to the following likelihood combinations:
\begin{itemize}
    \item Agnostic baseline + $(A_{\rm s}, n_{\rm s})$ priors
        \item Agnostic baseline + $(A_{\rm s}, n_{\rm s})$ priors + CMB lensing
    \item Standard baseline + $(A_{\rm s}, n_{\rm s})$ priors + CMB lensing
\end{itemize}
In Appendix \ref{app:asns}, we relax the priors on $A_{\rm s}$ and $n_{\rm s}$ to prove that their role is actually not crucial in our analysis.

All our runs assume a flat and uninformative prior on the standardized SNIa magnitude $M_B$ (for the SNIa likelihood). 
The runs with the agnostic baseline likelihood further assumes a flat uninformative prior on the unknown sound horizon $r_{\rm s}$.

\begin{table*}[!tbp]
\centering
\begin{adjustbox}{width=\textwidth}
\begin{tabular}{|c|cc|cc|cc|}
\hline

\multirow{2}{*}{Parameters}  & \multicolumn{2}{c|}{ Agnostic Baseline+$A_{\rm s}+n_{\rm s}$} & \multicolumn{2}{c|}{ Agnostic Baseline+$A_{\rm s}+n_{\rm s}$+CMB lensing} & \multicolumn{2}{c|}{Standard Baseline+$A_{\rm s}+n_{\rm s}$+CMB lensing} \\

\cline{2-7}
& $\Lambda$CDM & $w_0w_a$ & $\Lambda$CDM & $w_0w_a$ & $\Lambda$CDM & $w_0w_a$   \\
\hline \hline
{\boldmath$\omega_\mathrm{b}$} & $2.218_{-0.060}^{+0.060}$ & $2.218_{-0.055}^{+0.054}$ &$2.222_{-0.060}^{+0.061}$  &$2.219_{-0.060}^{+0.061}$ &$2.202_{-0.056}^{+0.052}$&$2.211_{-0.058}^{+0.053}$\\

{\boldmath$\omega_\mathrm{cdm}$} &$0.155_{-0.008}^{+0.025}$&$0.151_{-0.010}^{+0.030}$ &$0.156_{-0.010}^{+0.018}$ &$0.149_{-0.016}^{+0.016}$ & $0.1179_{-0.0009}^{+0.0010}$& $0.1186_{-0.0012}^{+0.0014}$\\

{\boldmath$H_0$ [km/s/Mpc]} &$79.1_{-3.3}^{+6.7}$ & $76.6_{-4.2}^{+7.4}$ &$79.3_{-3.1}^{+5.2}$ &$76.0_{-4.3}^{+4.5}$ &$68.23_{-0.48}^{+0.45}$ & $67.41^{+0.68}_{-0.68}$\\

{\boldmath$\ln 10^{10}A_{s}$} & $3.038_{-0.021}^{+0.021}$ &$3.038_{-0.021}^{+0.021}$ &$3.035_{-0.020}^{+0.019}$&$3.038_{-0.020}^{+0.021}$ &$3.055_{-0.016}^{+0.016}$ & $3.048_{-0.017}^{+0.017}$\\

{\boldmath$n_{\rm s}$} & $0.959_{-0.010}^{+0.010}$ & $0.959_{-0.010}^{+0.010}$ &$0.9587_{-0.0092}^{+0.0092}$ &$0.9581_{-0.0096}^{+0.0091}$ &$0.9576_{-0.0085}^{+0.0087}$& $0.9574_{-0.0092}^{+0.0088}$\\

{\boldmath$\sum m_\mathrm{\nu}$ [eV]} &$1.08_{-0.60}^{+0.65}$ &$1.00_{-0.99}^{+0.33}$ &$1.01_{-0.46}^{+0.52}$&  $0.83_{-0.54}^{+0.42}$& $<0.097$& $<0.15$\\

{\boldmath$w_0$} & $-$& $-0.870_{-0.067}^{+0.064}$ &$-$&$-0.875_{-0.063}^{+0.061}$ &$-$ & $-0.854_{-0.058}^{+0.055}$\\

{\boldmath$w_a$}  & $-$& $-0.39_{-0.28}^{+0.30}$ &$-$&$-0.36_{-0.26}^{+0.30}$ &$-$& $-0.50_{-0.20}^{+0.27}$\\
\hline

$\Omega_\mathrm{m}$ & $0.3007_{-0.0046}^{+0.0045}$ &$0.3108_{-0.0065}^{+0.0065}$ & $0.2998_{-0.0045}^{+0.0041}$&$0.3102_{-0.0065}^{+0.0063}$ &$0.3014_{-0.0043}^{+0.0040}$& $0.3111_{-0.0060}^{+0.0058}$\\

$S_8$ & $0.809_{-0.019}^{+0.022}$ & $0.809_{-0.019}^{+0.022}$ &$0.817_{-0.013}^{+0.015}$ &$0.820_{-0.013}^{+0.014}$ &$0.8155_{-0.0096}^{+0.0093}$&$0.822_{-0.010}^{+0.012}$ \\

$r_{\rm s}$ [Mpc] &$126_{-12}^{+3}$& $128_{-12}^{+6}$ & $125.9_{-8.2}^{+4.2}$ &$129.2_{-8.2}^{+6.7}$ &$145.36_{-0.57}^{+0.60}$ & $145.10_{-0.64}^{+0.61}$\\

\hline
$ \chi^2_{\rm min}$ & 1424.1& 1418.7&1435.0 & 1438.4&1447.4&1441.8\\
\hline
\end{tabular}
\end{adjustbox}
\caption{\label{tab:params_lcdm1} 
Mean values  and credible intervals (68\%CL) for cosmological and selected derived parameters of the $\Lambda$CDM or $w_0w_a$CDM model, fit to combinations of data described in the text, along with best-fit $\chi^2$ values. For neutrino masses and in the case of the standard analysis, we quote instead the upper 95\%CL of the credible interval.}
\end{table*}

\begin{figure*}
    \centering
    \includegraphics[width=0.8\linewidth]{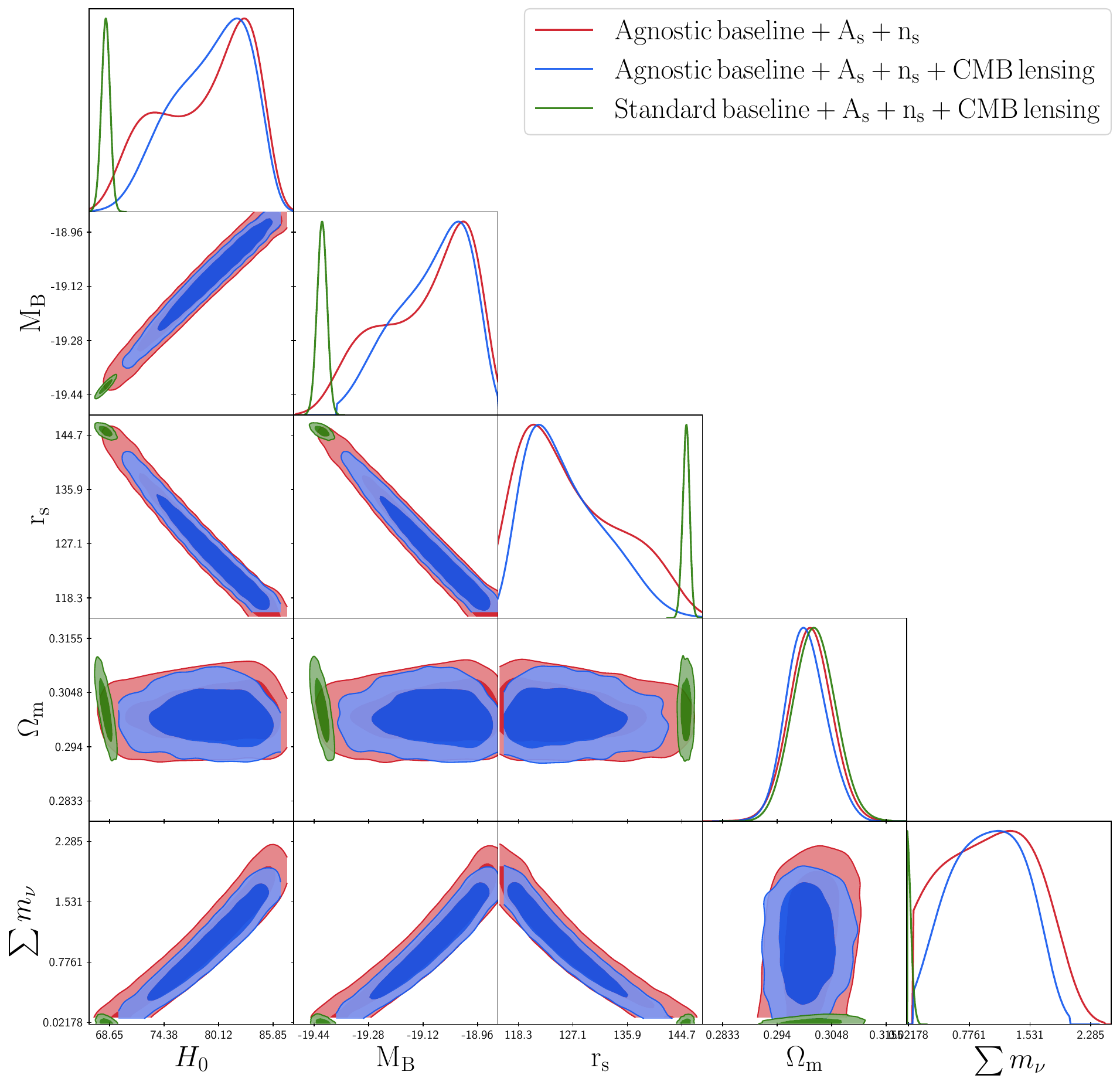}
    \caption{Assuming the $\Lambda$CDM+$\sum m_{\nu}$ model w/o a free sound horizon, triangle plot for ($H_0$, $M_B$, $r_{\rm s}$, $\Omega_{\rm m}$, $\sum m_\nu$) for different data sets (see legend and text), marginalized over ($\omega_{\rm b}$, $\omega_{\rm cdm}$, $A_{\rm s}$, $n_{\rm s}$).}
    \label{fig:traingle_lcdm}
\end{figure*}

\begin{figure*}
    \centering
    \includegraphics[width=0.8\linewidth]{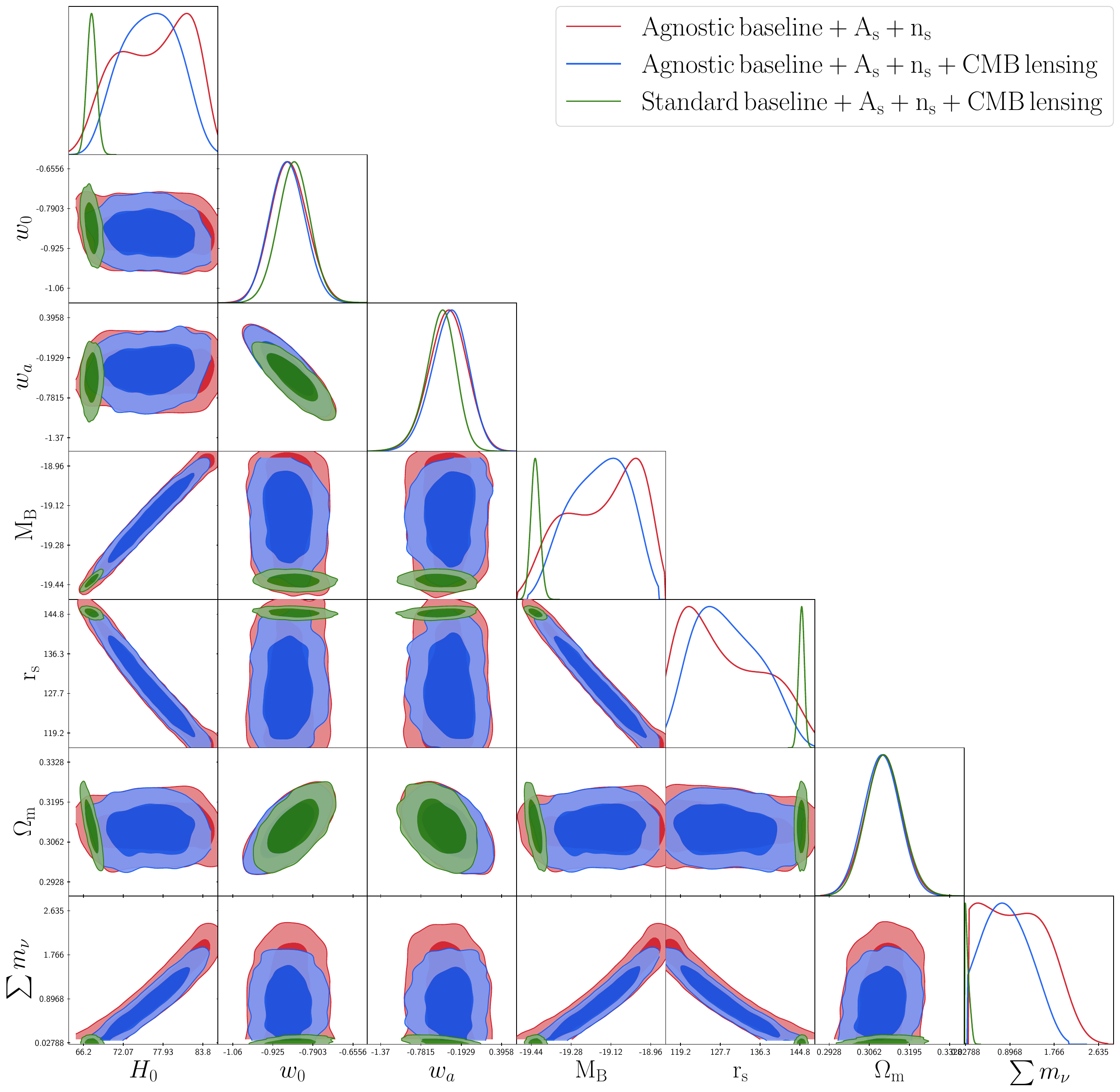}
   \caption{Assuming a $w_0 w_a$CDM+$\sum m_{\nu}$ model w/o a free sound horizon, triangle plot for ($H_0$, $w_0$, $w_a$, $M_B$, $r_{\rm s}$, $\Omega_{\rm m}$, $\sum m_\nu$) for different data sets (see legend and text), marginalized over ($\omega_{\rm b}$, $\omega_{\rm cdm}$, $A_{\rm s}$, $n_{\rm s}$).}
    \label{fig:traingle_w0wa}
\end{figure*}

Our results are presented in Table \ref{tab:params_lcdm1} and in the triangle plots of Figure~\ref{fig:traingle_lcdm} (assuming a cosmological constant) or Figure~\ref{fig:traingle_w0wa} (assuming CPL dark energy). Our most striking results can be summarized in a few points.\\

{\bf ``Standard baseline + $A_{\rm s}$ + $n_{\rm s}$ + CMB lensing'' case.}    
    In this work, we use this case as a point of comparison with respect to the agnostic runs. Therefore, it is useful to understand how it differs from a more traditional parameter inference based on full data from CMB, BAO, SNIa and cosmic shear observations. Technically, the difference lies in the replacement of data on the CMB temperature and polarization spectra (2-point statistics) with a set of priors on the four quantities ($\omega_{\rm b}$, $A_{\rm s}$, $n_{\rm s}$, $\theta_{\rm s}$).
    The results, reported in the last two columns of table \ref{tab:params_lcdm1}, show that most parameter constraints are similar in our ``Standard baseline + $A_{\rm s}$ + $n_{\rm s}$'' approach and in the traditional approach (reported, for instance, in table VII of \cite{DESI:2025ejh}, up to the inclusion or not of an $S_8$ prior). As expected, the ``Standard baseline + $A_{\rm s}$ + $n_{\rm s}$'' analysis prefers a low value of $H_0$ and has a 2.2$\sigma$ preference for CPL dark energy over a plain cosmological constant. Like with the traditional approach, we find that allowing for dynamical dark energy shifts the preferred range for the Hubble parameter to even smaller values. The main difference between the credible intervals derived from our ``Standard baseline + $A_{\rm s}$ + $n_{\rm s}$'' or from the traditional approach concerns the neutrino mass. The reason is that the high-$\ell$ CMB temperature/polarization spectra contain additional (although model-dependent) information on CMB lensing compared to the reconstructed lensing potential spectrum. Already in the $\Lambda$CDM+$\sum m_\nu$ case, the ``Standard baseline + $A_{\rm s}$ + $n_{\rm s}$'' data set is compatible with neutrino masses not in tension with oscillation data (assuming NH), $\sum m_{\nu}\le 0.0965$~eV (95\%CL), and even peaking at a non-zero value. In the $w_0 w_a$CDM+$\sum m_\nu$ case, the confidence interval widens to $m_{\nu}\le 0.147$~eV (95\%CL). This shows that, in traditional analyses using full-shape CMB temperature and polarization, the CMB lensing information contained in the 2-point statistics plays a really important role compared to that extracted from the 4-point statistics. However, the former strongly depends on the assumed spectrum of primary CMB anisotropies, while the later is marginalized over this information and only probes the growth of structures. In this sense, the neutrino mass constraints derived from our ``Standard baseline + $A_{\rm s}$ + $n_{\rm s}$'' data set are already quite robust and conservative, even if they still rely on standard predictions for the sound horizon.\\

{\bf ``Agnostic baseline + $A_{\rm s}$ + $n_{\rm s}$ + CMB lensing'' case.}    
    In this case, following the discussion of section \ref{sec:preliminary}, we expect that the uncalibrated IDL and SNIa data constrain the dark energy parameters ($\Omega_{\rm m}=1-\Omega_\Lambda$ and potentially $w_0$, $w_a$) and correlate $H_0$ with $r_{\rm s}$. We further expect that the combination of priors on the primordial spectrum, BBN and the growth of structures constrain $H_0$. The results are displayed in figures \ref{fig:traingle_lcdm} and \ref{fig:traingle_w0wa}. The confidence interval for the Hubble parameter is now  $H_0=79.3^{+5.2}_{-3.1} [\mathrm{km/s/Mpc}]$ (68\%CL) when assuming a cosmological constant, or $H_0=76.0^{+4.5}_{-4.4} [\mathrm{km/s/Mpc}]$ (68\%CL) with dynamical dark energy. In both cases, the data is now well compatible with the high value of $M_B$ measured by SH0ES, and even with much larger values: we confirm that models with a non-standard sound horizon have the full potential to solve the Hubble tension. The values of $H_0$ and $r_{\rm s}$ required by the IDL data with a standard sound horizon are at the edge of the 2$\sigma$ posteriors. Neutrino mass bounds are also relaxed in the agnostic analysis, since $\sum m_\nu$ correlates with $H_0$ and $r_{\rm s}$ -- as a consequence of its impact on the angular distance $d_A(z_*)$. The sign of the correlation between $\sum m_\nu$ and $H_0$ is reversed compared to the standard analysis, because this correlation now comes mainly from information on structure growth rather than distances: in an older universe, structures grow more, unless this gets compensated by more neutrino free-streaming. The agnostic results allow for $\sum m_\nu =1.01_{-0.46}^{+0.52}$~eV (68\% CL) assuming a cosmological constant, or $\sum m_\nu =0.830^{+0.42}_{-0.54}$~eV (68\% CL) assuming dynamical dark energy. Finally, we find that the parameters related to dark energy, ($\Omega_{\rm m}$, $w_0$, $w_a$), exhibit no correlation with $r_{\rm s}$ and are constrained identically in the standard and agnostic analyses. This shows that the information from uncalibrated BAO and SNIa data is sufficient to constrain the evolution of the dark energy component relative to the matter component, regardless of any prior on the calibration parameters $r_{\rm s}$ and $M_B$. In particular, in the agnostic analysis, we still find a preference for evolving dark energy over $\Lambda$ at the $2.2\sigma$ level, as can be seen in figure \ref{fig:traingle_w0wa}. \\
    %

    {\bf ``Agnostic baseline + $A_{\rm s}$ + $n_{\rm s}$'' case.}
    We can get even more conservative results by removing CMB lensing data from our analysis. In this case, also shown in figures \ref{fig:traingle_lcdm} and \ref{fig:traingle_w0wa}, the parameters $H_0$, $r_{\rm s}$ and $\sum m_\nu$ are only constrained by a combination of priors on $A_{\rm s}$, $n_{\rm s}$, $\omega_{\rm b}$, and $S_8$ (inferred from small-$\ell$ CMB data, BBN and cosmic shear). The posteriors on these parameters are relaxed in the absence of CMB lensing, but not dramatically -- only by 10\% to 20\%. Thus, the conclusion that the agnostic analysis prefers large $H_0$ values and is compatible with large neutrino masses is not driven specifically by CMB lensing data: it also hold when $H_0$ and $\sum m_\nu$ are only bounded by a prior on $S_8$.\\ 
 
    {\bf Removing priors on the primordial spectrum.}
    We justified the use of $A_{\rm s}$ and $n_{\rm s}$ priors saying that they are constrained by large-scale CMB data, which are independent of the sound horizon. One may argue that our $n_{\rm s}$ prior is not conservative, since it is based on fits to the whole CMB spectra. However, we show in appendix \ref{app:asns}  that the results of the ``agnostic baseline + $A_{\rm s}$ + CMB lensing'' and ``agnostic baseline + $A_{\rm s}$ + $n_{\rm s}$ + CMB lensing'' analyses are nearly identical, which proves that the $n_{\rm s}$ prior plays a negligible role in our results. Appendix \ref{app:asns} actually shows that even without an $A_{\rm s}$ prior, the ``agnostic baseline + CMB lensing'' data set is sufficient to constrain the five parameters ($H_0$, $M_B$, $r_{\rm s}$, $\Omega_{\rm m}$, $\sum m_\nu$), thanks to the shape information in the CMB lensing data. Thus, our main results are robust not only against CMB lensing data, but also against priors on the primordial spectrum. In connection to this point, we note that in our analysis, the determination of $\tau_{\rm reio}$ does not play a direct role, but could be indirectly relevant. Indeed, if $\tau_{\rm reio}$ was different from the value measured by {\it Planck} (using the low-$\ell$ peak in the polarization spectrum), our $A_{\rm s}$ prior would need to be shifted, since the CMB unambiguously constrains the combination $e^{-2\tau_{\rm reio}}A_{\rm s}$. But since the results of the agnostic analysis are nearly independent of the $A_s$ prior, they are also independent of $\tau_{\rm reio}$. This is not the case with standard assumption on the sound horizon, as pointed out in \cite{Sailer:2025lxj,Jhaveri:2025neg}.\\

{\bf Comparison with previous work.} Our results in the $\Lambda$CDM+$\sum m_\nu$ agnostic case are very well compatible with those of \cite{GarciaEscudero:2025lef}, despite slightly different choices for the CMB lensing data and for priors on $\omega_{\rm b}$, $A_{\rm s}$, $n_{\rm s}$, $\theta_{\rm s}$. Besides, our work generalizes the result of  \cite{GarciaEscudero:2025lef} to the $w_0 w_a$CDM+$\sum m_\nu$ case. It also provides various comparisons allowing to identify the role of each individual data set in the final results.

\section{Discussion and conclusions}
\label{sec:discussion}

The most interesting results of this analysis can be summarized in three points regarding either dark energy, the Hubble parameter or the summed neutrino mass. \\

{\bf Dark energy. } We find that {\it the dark energy evolution is well constrained by uncalibrated IDL data and SNIa data, irrespectively of any prior on the sound horizon.} We recall that ``uncalibrated IDL data'' refers to information on BAO and CMB peaks marginalized over the unknown value of the sound horizon. We also recall that in our case, the parameters describing the evolution of dark energy relative to matter are $\Omega_{\rm m}$ (or equivalently $\Omega_\Lambda$), $w_0$ and $w_a$. The posteriors on these parameters remain nearly identical when $r_{\rm s}$ is predicted by standard cosmology or freely varied and marginalized.  The explanation is that the range of redshifts over which the universe might be DE-dominated is very well covered by BAO and SNIa data. Knowing only the uncalibrated redshift-dependence of $d_L$ and $d_A$ is sufficient to probe the shape of the expansion history $H(z)$ up to its overall calibration, that is, for determining the DE parameters but not $H_0$. At first sight, it sounds like this conclusion is not aligned with that from references \cite{Chaussidon:2025npr} and \cite{Mirpoorian:2025rfp}, which show that specific models of EDE or modified recombination reduce the preference for (late) dynamical dark energy. Our interpretation is that the results of \cite{Chaussidon:2025npr,Mirpoorian:2025rfp} do not originate from a generic degeneracy at the level of geometrical information and distance measurements, since $\Omega_{\rm m}$ and $r_{\rm s}$ are not correlated when only geometrical data is involved. Instead, the results of \cite{Chaussidon:2025npr,Mirpoorian:2025rfp} are likely to be induced by subtle parameter correlations when these specific models are fitted to the full shape of the CMB spectrum.  \\

{\bf Hubble rate.} We confirm the well-known expectation that fluctuating the sound horizon relaxes bounds on $H_0$. Additionally, we confirm previous results by \cite{Pogosian:2020ded,GarciaEscudero:2025lef} indicating that {\it in an analysis agnostic to the sound horizon and in presence of data on the growth of structure, very large values of $H_0$ are preferred, perfectly consistent with SH0ES, but in mild tension with the low values indicated by the standard calibrated IDL method.} In the $\Lambda$CDM+$\sum m_\nu$ case, we find a 3.5$\sigma$ tension on $H_0$ between our standard and agnostic runs. In the $w_0w_a$CDM+$\sum m_\nu$ case, the two $H_0$ posteriors marginally agree at to the 2$\sigma$ level. Since the agnostic runs are well compatible with SH0ES measurements, we can combine the corresponding data set with the Gaussian prior on $M_B$ from Eq.~(\ref{eq:SHOES}). In the agnostic $\Lambda$CDM+$\sum m_\nu$ case, the joint fit to the ``agnostic baseline + $A_{\rm s}$ + $n_{\rm s}$ + CMB lensing + SH0ES'' returns
\begin{eqnarray}
r_{\rm s}&=&135.1 \pm 1.9~\mathrm{Mpc}~(68\%\mathrm{CL}),\nonumber \\
H_0&=&73.65 \pm 0.94~\mathrm{km/s/Mpc}~(68\%\mathrm{CL}),
\end{eqnarray}
to be compared with the standard model prediction $r_{\rm s}=145.4 \pm 0.6$~Mpc. In the agnostic $w_0 w_a$CDM+$\sum m_\nu$ case, we get similar results with a  SH0ES prior,
\begin{eqnarray}
r_{\rm s}&=&134.0 \pm 2.0~\mathrm{Mpc}~(68\%\mathrm{CL}),\nonumber \\
H_0&=&73.0 \pm 1.0~\mathrm{km/s/Mpc}~(68\%\mathrm{CL}).
\end{eqnarray}

{\bf Neutrino mass.} Our results indicate that {\it the cosmological neutrino mass bound can be significantly relaxed just by discarding information from high-$\ell$ CMB data, and even much more by being agnostic about the sound horizon.} Current neutrino mass bounds are known to be dominated by two effects of $\sum m_\nu$ on cosmological observables: its effect on geometry, and its effect on CMB lensing. In reality, this means that bounds on $\sum m_\nu$ may be dominated by three types of observational evidences: measurements of the angular distance to recombination, measurements of the lensing potential spectrum extracted from 4-point correlations of CMB maps, and bounds on CMB lensing inferred from 2-point correlations of CMB maps. Our analysis allows to test separately the impact of each of these three clues. The loose bounds on $\sum m_\nu$ found in the ``standard baseline + $A_{\rm s}$ + $n_{\rm s}$'' analysis show that the information contained in the 2-point correlation of CMB maps is currently playing a very important role, and actually drives the small tension between cosmological and laboratory bounds on neutrino masses. Without such information, we get:
\begin{equation}
\sum m_\nu \le 0.097~{\rm eV}
~~\mathrm{(}\Lambda\mathrm{CDM, 95\%CL)}~, 
\end{equation}
which is well compatible with the NH scenario and marginally compatible with the IH one. With dynamical dark energy, we get
\begin{equation}
\sum m_\nu \le 0.15~{\rm eV}
~~\mathrm{(}w_0 w_a\mathrm{CDM, 95\%CL)}~, 
\label{eq:bound_standard}
\end{equation}
which is perfectly compatible with both scenarios. It is worth stressing that the CMB lensing information derived from 2-point statistics does depend heavily on assumptions regarding the early cosmological evolution, because it comes from a fit to the spectrum of primary CMB anisotropies modulated by a second-order effect (CMB lensing). Instead, the lensing information derived from the 4-point correlation function only probes the CMB lensing potential, and as such, the large-scale structure of the universe, independently of what is assumed for primary CMB anisotropies. The even looser bounds on $\sum m_\nu$ found in the ``agnostic baseline + $A_{\rm s}$ + $n_{\rm s}$'' analysis show what happens when we further remove any constraint from geometry on $\sum m_\nu$. The bounds obtained in this case, 
\begin{align}
\sum m_\nu & =1.01_{-0.46}^{+0.52}~{\rm eV} & {\rm (}\Lambda{\rm CDM, 68\%CL)}~, \nonumber \\
\sum m_\nu &=0.83^{+0.42}_{-0.54} ~{\rm eV} &{\rm (}w_0w_a{\rm CDM, 68\%CL)}~,
\end{align}
only come from the information on  CMB lensing derived from the 4-point correlation of CMB map, complemented by extra information on the growth of structure contained in our $S_8$ prior. These bounds are compatible with high values of $\sum m_\nu$ potentially in tension with laboratory bounds on $\beta$-decay from KATRIN \cite{KATRIN:2024cdt}. They are also incompatible with data on neutrinoless double-$\beta$ decay from KamLAND-Zen \cite{KamLAND-Zen:2024eml} and GERDA \cite{GERDA:2020xhi}, but only in the case where neutrinos are Majorana particles. Thus, instead of assuming a flat prior on any positive value of $\sum m_\nu$, we can add the KATRIN prior described at the end or section \ref{sec:data}, and check whether we obtain a stronger upper bound on $H_0$ and a stronger lower bound on $r_{\rm s}$. With an additional KATRIN prior, we find that our main results are essentially unchanged in the agnostic $\Lambda$CDM case, while in the agnostic $w_0 w_a$CDM case the bounds narrow down to
\begin{align}
\sum m_\nu &= 0.69^{+0.33}_{-0.47} ~{\rm eV} &{\rm (68\%CL)}~, \nonumber \\
r_{\rm s}&= 131.1^{+6.8}_{-6.9}~\mathrm{Mpc}~&(68\%\mathrm{CL}),\nonumber \\
H_0&=74.7^{+3.4}_{-4.4}~\mathrm{km/s/Mpc} &(68\%\mathrm{CL}),
\end{align}
using the combination ``agnostic baseline + $A_{\rm s}$ + $n_{\rm s}$ + CMB lensing + KATRIN'' (but not SH0ES). These conservative bounds are compatible with cosmology and laboratory constraints that do not rely on a particular model for the calculation of the sound horizon, and constitute the main results of our work. 

\begin{acknowledgments}
We thank H. Garc\'{\i}a Escudero, S. H. Mirpoorian and L. Pogosian for illuminating discussions on an earlier version of this draft and for stressing the importance of the  \texttt{lens\_only} flag in the lensing likelihood for our type of analysis.
Computational work was performed with computing resources granted by RWTH Aachen University under project ‘rwth1661’ and 'p0021792'. RKS thanks the Alexander von Humboldt Foundation for their support. 
\end{acknowledgments}

\bibliographystyle{apsrev4-1}
\bibliography{output}

\appendix

\section{Comparison of Method I vs Method II}
\label{app:methods}

\begin{figure}
    \centering
    \includegraphics[width=0.8\linewidth]{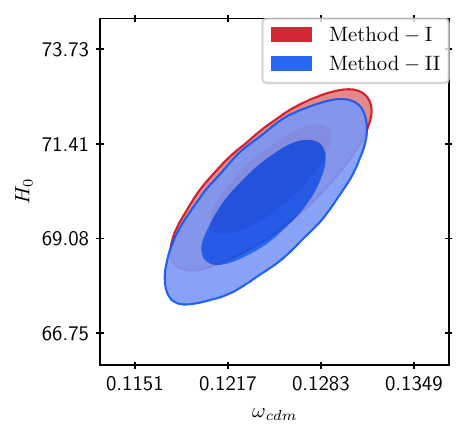}
    \caption{Assuming the $\Lambda$CDM  model with a free sound horizon but fixed values of ($A_{\rm s}$, $n_{\rm s}$, $\omega_{\rm b}$, $\sum m_\nu$), 2D contours on the free parameters $(H_0, \omega_{\rm ncdm})$ inferred from the uncalibrated IDL likelihood, implemented either with Method I or Method II (see section \ref{sec:agnostic}). In both cases, $r_{\rm s}$ is floated freely and marginalised over (either implicitly or explicitly).}
    \label{fig:app_methods}
\end{figure}

As discussed in main text in section \ref{sec:AgnosticEP}, we can proceed in two ways to be agnostic to the sound horizon, either by taking a standard version of the likelihoods and making $r_{\rm s}$ a free parameter (Method II), or by taking ratios of BAO and CMB observables that no longer depend on the sound horizon (Method I). The only small difference between these two approaches is that they assume different quantities to be gaussian-distributed (either distances or ratios of distances). Thus, they contain the same information, but they are not fully equivalent in terms of priors. We confirm this through dedicated MCMC runs. Like in section \ref{sec:preliminary}, we fix ($A_{\rm s}$, $n_{\rm s}$, $\omega_{\rm b}$, $\sum m_\nu$), we stick to a plain cosmological constant, and we only vary $(H_0, \omega_{\rm ncdm})$ -- as well as $r_{\rm s}$ in Method II. Our results for the $(H_0, \omega_{\rm ncdm})$ contours are presented in Figure \ref{fig:app_methods}. Despite their slightly different priors, the two approaches give almost the same results. 

\section{Effect of priors on primordial spectra parameters ($A_{\rm s},n_{\rm s}$)\label{app:asns}}

\begin{figure*}
    \centering  \includegraphics[width=0.8\linewidth]{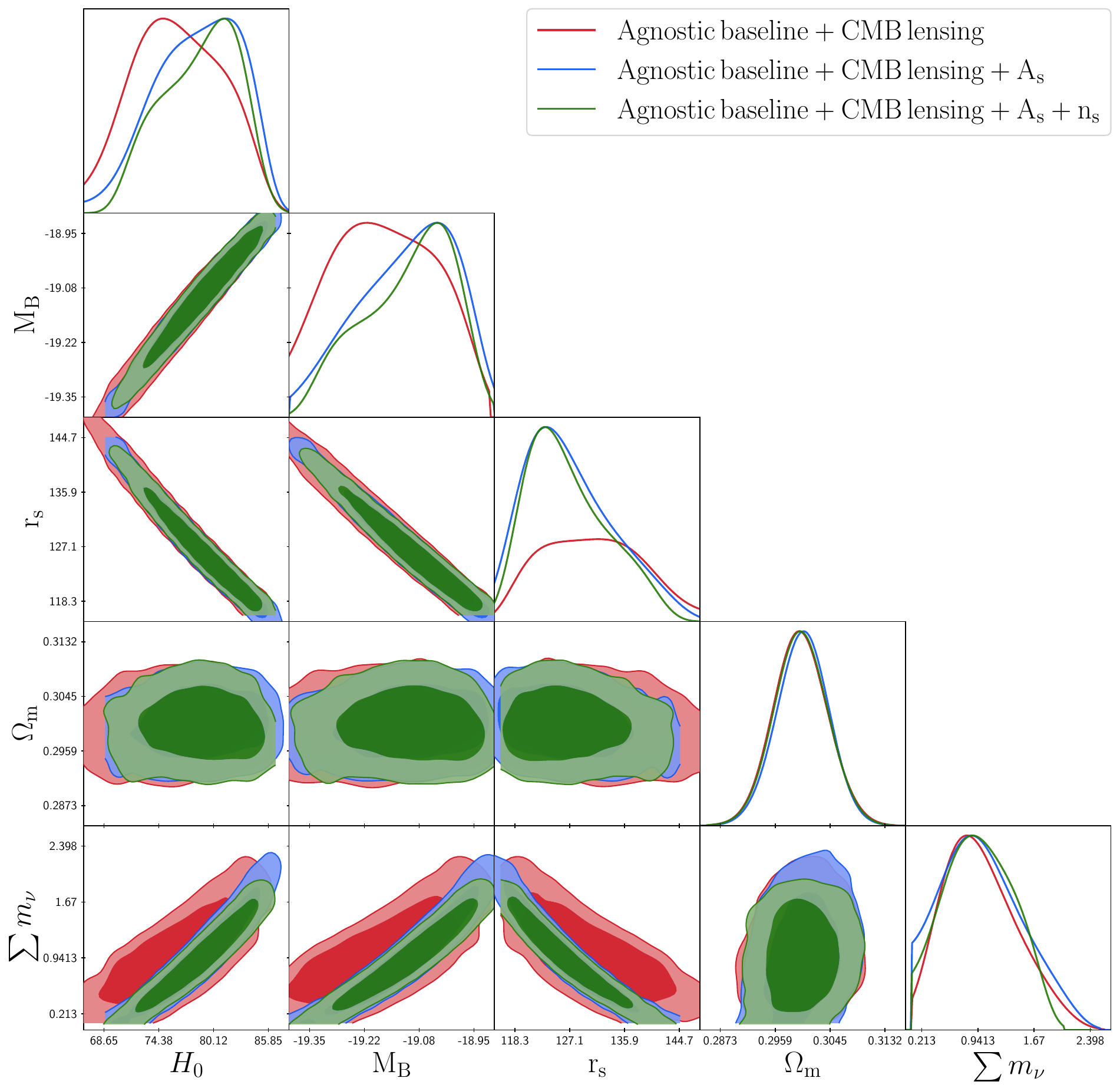}
    \caption{Assuming a $\Lambda$CDM+$\sum m_{\nu}$ model w/o a free sound horizon, triangle plot for ($H_0$, $M_B$, $r_{\rm s}$, $\Omega_{\rm m}$, $\sum m_\nu$) for different data sets (see legend and text), marginalized over ($\omega_{\rm b}$, $\omega_{\rm cdm}$, $A_{\rm s}$, $n_{\rm s}$).}
    \label{fig:as_ns_lcdm}
\end{figure*}

\begin{figure*}
    \centering
\includegraphics[width=0.8\linewidth]{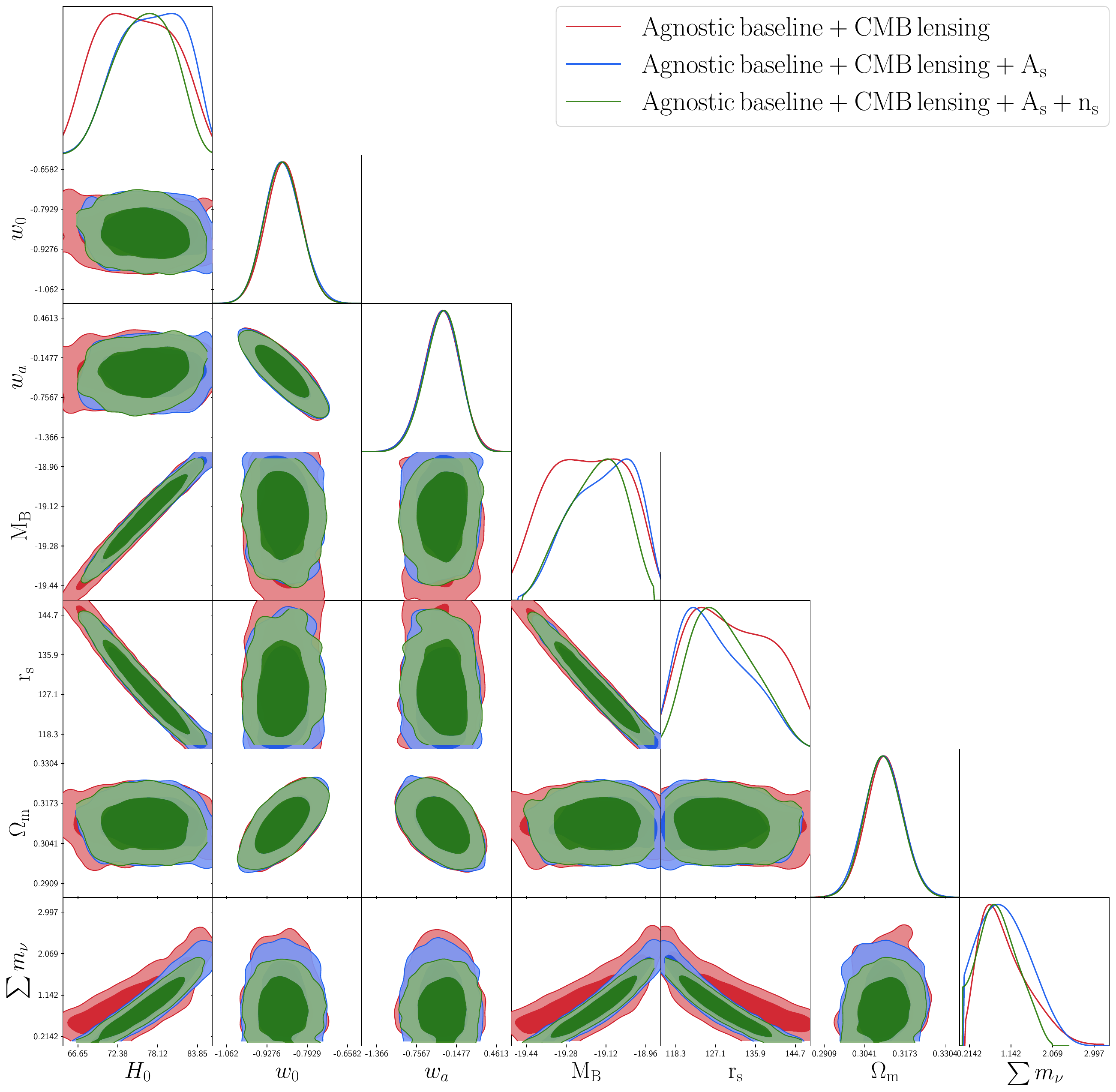}
    \caption{Assuming a $w_0 w_a$CDM+$\sum m_{\nu}$ model w/o a free sound horizon, triangle plot for ($H_0$, $w_0$, $w_a$, $M_B$, $r_{\rm s}$, $\Omega_{\rm m}$, $\sum m_\nu$) for different data sets (see legend and text), marginalized over ($\omega_{\rm b}$, $\omega_{\rm cdm}$, $A_{\rm s}$, $n_{\rm s}$).}
    \label{fig:as_ns_w0wa}
\end{figure*}

Our main results correspond to the ``agnostic baseline + $A_{\rm s}$ + $n_{\rm s}$ + CMB lensing'' data set. In section \ref{sec:results}, we already showed how the results relax when removing CMB lensing data. Here, we keep CMB lensing but we remove the priors on the primordial spectrum (either the $n_{\rm s}$ prior, or both priors on $A_{\rm s}$ and $n_{\rm s}$). The results are presented in figure \ref{fig:as_ns_lcdm} for $\Lambda$CDM+$\sum m_\nu$ and in figure \ref{fig:as_ns_w0wa} for $w_0 w_a$CDM+$\sum m_\nu$. In all these cases, the bounds on ($H_0$, $M_B$, $r_{\rm s}$, $\Omega_{\rm m}$, $\sum m_\nu$) remain very stable. This shows that priors on the primordial spectrum do not play an important role in our analysis as long as we retain data on the full shape of CMB lensing. As explained in section \ref{sec:results}, this implies that assumptions on the value of $\tau_{\rm reio}$ are not playing a significant role either, not even indirectly.

\end{document}